\documentclass[aps,showpacs,twocolumn]{revtex4}
\usepackage{amsfonts}
\usepackage{amsmath}
\usepackage{amssymb}
\usepackage{graphicx}
\usepackage{bm}
\usepackage{color}
\usepackage{subfigure}
  
\graphicspath{{C:/Users/a.ly/Desktop/image/}}
\begin{document}

\title{Hydrodynamic interactions in DNA\ thermophoresis}. 
\author{Aboubakry Ly and Alois W\"{u}rger}
\affiliation{Laboratoire Ondes et Mati\`{e}re d'Aquitaine, Universit\'{e} de Bordeaux \&
CNRS, 33405 Talence, France}

\begin{abstract}
We theoretically study the molecular-weight dependence of DNA thermophoresis, 
which arises from mutual advection of the $n$ repeat units of the molecular chain. As a main result 
we find that the dominant driving forces, i.e., the thermally induced permittivity gradient 
and the electrolyte Seebeck effect, result in characteristic hydrodynamic screening. 
In comparison with recent experimental data on single-stranded DNA ($2\le n\le80$), our 
theory provides a good description for the increase of the drift velocity up to $n=30$;
the slowing-down of longer molecules is well accounted for by a simple model for counterion 
condensation. It turns out that thermophoresis may change sign as a function of $n$: For an 
appropriate choice of the salt-specific Seebeck coefficient, short molecules move to the cold 
and long ones to the hot; this could be used for separating DNA by molecular weight. 
\end{abstract}

\maketitle

When applying a temperature gradient on a colloidal dispersion, one observes
thermally driven transport towards the hot or the cold 
\cite{Pia08,Wue10}.  In recent years, thermophoresis has been shown
to provide a versatile means for manipulating DNA, including translocation
through plasmonic nanopores \cite{Nic14}, stretching in nanochannels \cite%
{He13,Ped14}, separation by molecular weight \cite{Mae11}, sequence-specific 
detection with functionalized nanoparticles \cite{Yu15}, and force-free trapping 
of single molecules \cite{Bra16}. Protein thermophoresis has become a standard 
technology in biomedical analysis \cite{Jer14}, and the accumulation of RNA in 
hydrothermal pores is discussed as a scenario for biomolecular synthesis in the 
early evolution of life \cite{Baa07}.

In the last decade, much progress has been made concerning the physical
mechanisms of thermophoresis of charged colloids. It has been shown that, 
in addition to thermo-osmosis \cite{Ruc81,Bre16}, the electrolyte Seebeck field 
\cite{Put07,Wue08,Vig10,Esl14,Rei14,Sim16} and concentration gradients of
salt \cite{Esl14} or nonionic polymers \cite{Jia09,Tsu17}, play an important role. 
These companion fields arise from specific solvation enthalpies of salt ions or 
nonionic solutes, and are at the origin of the \textquotedblleft inverse\textquotedblright\ 
Soret effect, where the colloids accumulate in hot regions \cite{Esl14,Jia09}.
Regarding the size dependence, there is conclusive evidence that
the mobility of colloidal beads does not vary with the radius \cite%
{Bra08,Wue16}. 

In spite of the many experimental studies mentioned above, little is known 
on the molecular-weight dependence of DNA thermophoresis. If the  
hydrodynamic slowing-down of Brownian motion is well understood 
in terms of mutual advection of the repeat units \cite{Gen79}, a more 
complex picture arises for phoretic motion where external forces are absent 
and which is driven by non-equilibrium surface properties. 
For short-ranged dispersion forces, hydrodynamic interactions 
are irrelevant and the thermophoretic velocity is constant \cite{Bro81,Wie04};
deviations observed for very short polymers in organic solvents \cite{Sta08}, 
arise probably from chemically different end groups. For DNA in a weak electrolyte, 
however, the electrostatic interaction length may attain tens of nanometers, 
which suggests an incomplete screening of hydrodynamic coupling.

In this paper we study hydrodynamic effects on DNA thermophoresis. We consider the 
two dominant charge-related surface forces, i.e., the thermally induced permittivity 
gradient and the electrolyte Seebeck effect, and derive the respective 
hydrodynamic correction factors in the rigid-rod limit. With a simple model 
for counterion condensation, we compare our theory to recent Soret data 
for single-stranded DNA \cite{Rei14}.

\begin{figure}[tbp]
\includegraphics[width=\columnwidth]{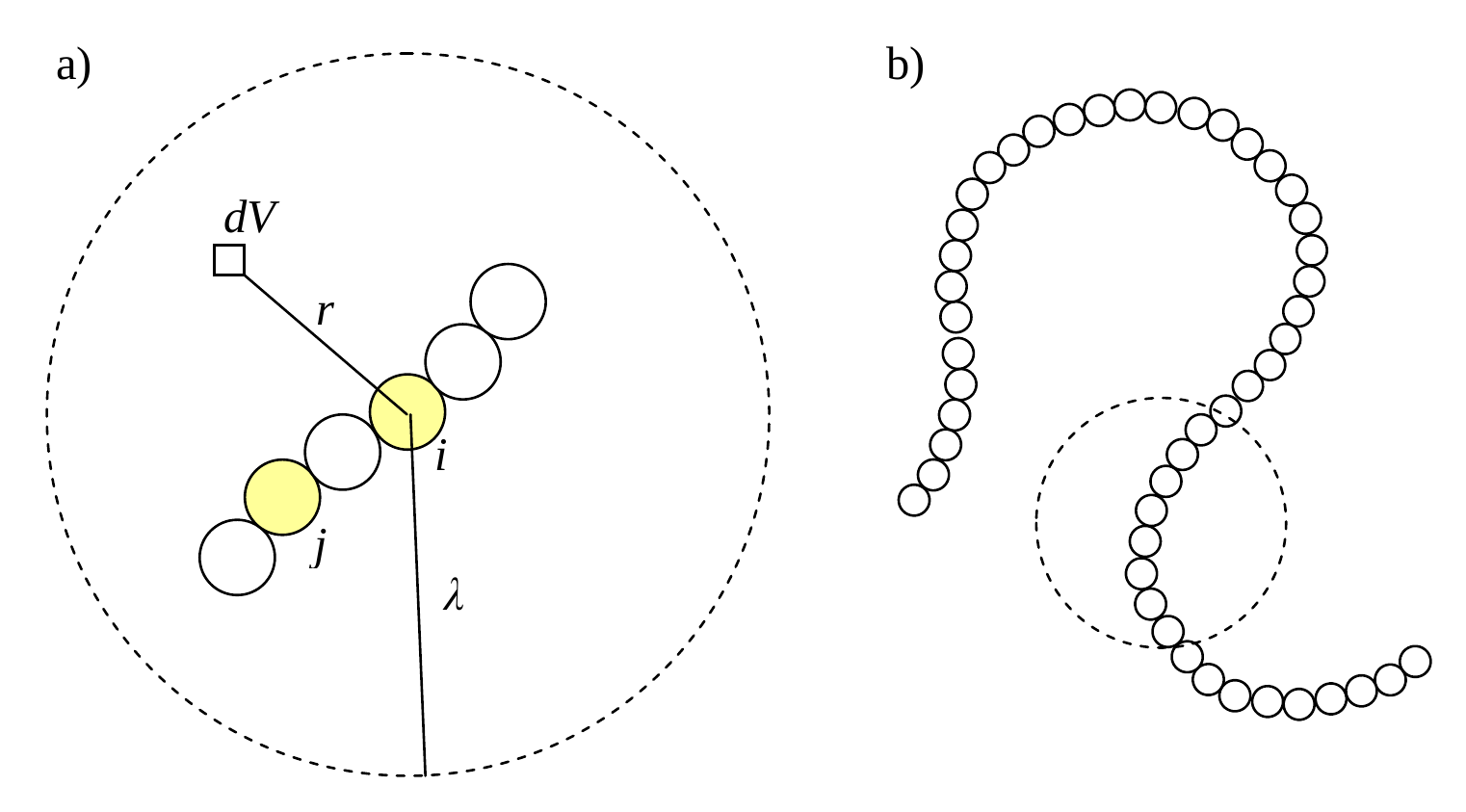}
\caption{Schematic view of a charged polymer in an electrolyte solution.\ a)
The drag on the molecular unit $j$ consists of two contributions: First, the stokeslet 
of bead $i$, driven by the force $\mathbf{F}$, drags the neighbor $j$. Second, 
due to the force $\mathbf{f}(\mathbf{r})$ exerted by the bead $i$, the fluid
element $dV$ moves and in turn exerts a drag on bead $j$.\ These
contributions cancel each other at distances well beyond the Debye length 
$\protect\lambda $. b) Since the persistence length is of the order 
of the Debye length, the molecule may be treated as a rigid rod within the reach 
of electrostatic interactions.}
\label{fig3}
\end{figure}

{\it Hydrodynamic interactions. --} 
Consider a polyelectrolyte chain of $n$ building blocks, as illustrated in Fig. \ref{fig3}. 
Unit $i$ creates a flow field $\mathbf{v}(\mathbf{r}-\mathbf{r}_i)$ in 
the surrounding fluid and thus drags its neighbor $j$. Then the overall velocity
$\mathbf{u}$ of the chain is given by the sum of the monomer contribution 
$\mathbf{u}_1$ and the mutual  advection,
\begin{equation}
\mathbf{u}=\mathbf{u}_{1}+\frac{1}{n}\sum_{i,j\neq i}\left\langle 
      \mathbf{v}(\mathbf{r}_{ij})\right\rangle , 
 \label{eq3}
\end{equation}%
where the angular brackets $\left\langle \cdot \cdot \cdot \right\rangle $
indicates the configurational average with respect to 
$\mathbf{r}_{ij}=\mathbf{r}_{j}-\mathbf{r}_{i}$. 

Thermophoresis arises from the  solute-solvent interactions. The force density exerted 
on the counterion cloud surrounding a charged monomer reads as \cite{Mor08}
\begin{equation}
\mathbf{f} = -\frac{E^{2}}{2}\mathbf{\nabla }\epsilon + \rho \mathbf{E}_{T},
\label{eq5}
\end{equation}%
where the first term is proportional to the thermally induced permittivity gradient 
$\mathbf{\nabla }\epsilon=(d\epsilon/dT)\mathbf{\nabla }T$, with the charged monomer's 
electric field $E$. Since the permittivity decreases with rising temperature,  $d\epsilon/dT<0$, 
the surrounding water moves to the hot, as recently confirmed experimentally for 
thermoosmosis in a capillary \cite{Bre16}. By reaction, the molecule migrates toward the cold. 

The second term in (\ref{eq5}) describes the force exerted by the macroscopic 
thermoelectric field $\mathbf{E}_T= S\mathbf{\nabla }T$  on the monomer's counterion 
density $\rho$ \cite{Wue08}. The electrolyte  Seebeck coefficient $S$ is a salt-specific 
quantity that may take either sign, resulting in motion along the temperature gradient or 
opposite to it \cite{Esl14}. Eq. (\ref{eq5}) gives the dominant thermal forces to leading 
order in the ratio $a/\lambda$ of the monomer radius and the Debye length. 
Additional companion fields, such as the salinity gradient, arise in the colloid limit  where 
$a$ is comparable to or larger than $\lambda$ \cite{Esl14}.

The force density $\mathbf{f}$ acts on the surrounding water and, by reaction, the molecular 
unit is subject to the opposite force $\mathbf{F}=-\int dV\mathbf{f}$ \cite{Mor08}. Thus the 
velocity field induced by the moving bead $i$ at the position of its neighbor $j$, consists of two 
contributions, 
\begin{equation}
\mathbf{v}(\mathbf{r}_{ij})=  \mathbf{G}(\mathbf{r}_{ij})\cdot \mathbf{F} 
      + \int \mathbf{G}(\mathbf{r}_{ij}-\mathbf{r})\cdot \mathbf{f}(\mathbf{r}) dV,  
\label{eq4}
\end{equation}
where $\mathbf{G}(\mathbf{r})=(1+\widehat{\mathbf{r}}\widehat{\mathbf{r}})/8\pi \eta r$ is 
the Oseen tensor with the viscosity $\eta$ and $\widehat{\mathbf{r}}=\mathbf{r}/r$ \cite{Kim91}.
The first term describes the long-range velocity field $v\sim 1/r_{ij}$ or ``stokeslet'' of 
 particle $i$ at the position $j$, due to the force $\mathbf{F}$; it gives rise to strong hydrodynamic 
effects on diffusion and sedimentation \cite{Gen79}. The second term is characteristic for phoretic 
motion;  it may be viewed as the sum of stokeslet flows of strength $\mathbf{f} dV$ and centered 
at a distance $\mathbf{r}$ from particle $i$, as illustrated in Fig. 1a. Since  both $E$ and $\rho$ 
vanish well beyond the Debye length, the second term cancels the first one at large distances, 
$r_{ij}\gg \lambda$, whereas it is small for nearby beads. 

As a consequence of this hydrodynamic screening, the advection velocity (\ref{eq4}) varies as 
$1/r_{ij}$ within the Debye length but vanishes at larger distances. When performing the 
configurational average in (\ref{eq3}) with the (isotropic) equilibrium distribution function, the 
only finite component of the mean drag velocity is along the force density $\mathbf{f}$,  that is,
along the temperature gradient. Then the tensor equation  simplifies to a scalar one, and
Eq. (\ref{eq3}) becomes 
\begin{equation}
u=u_{1}+\frac{1}{n}\sum_{i,j\neq i} \int \big< 
  G(\left\vert\mathbf{r}_{ij}-\mathbf{r}\right\vert ) - G(r_{ij}) \big> f(\mathbf{r}) dV ,
\label{eq7}
\end{equation}
with the Oseen tensor replaced by its diagonal part  $G(r)=1/6\pi \eta r$. This 
form shows that mutual advection vanishes for distant pairs with $r_{ij}\gg\lambda$, thus 
nicely displaying hydrodynamic screening.

In order to evaluate  (\ref{eq7}) we need to explicit the force density $f(\mathbf{r})$. 
The electrostatic potential of a single bead of valency $\hat{z}$ is well described by 
the  Debye-H\"{u}ckel expression 
\begin{equation}
\psi = - \frac{\hat{z} e}{4\pi \epsilon r}e^{-r/\lambda } = \zeta_1 \frac{a}{r}e^{-r/\lambda }, 
 \label{eq9}
\end{equation}
where the second equality defines the single-bead surface potential 
$\zeta_1=-\hat{z}e/4\pi \epsilon a$, which we assume to be negative. 
One readily  obtains the radial electric field $E=-d\psi/dr$ and the counterion charge density 
$\rho=-\epsilon \psi/\lambda^2$ which determine the force density (\ref{eq5}).
Then the volume integrals in (\ref{eq7}) can be performed in closed form \cite{Supp}, resulting 
in the factors $\left\langle e^{-2r_{ij}/\lambda}/r_{ij}^{2}\right\rangle$ and  
$\left\langle e^{-r_{ij}/\lambda}/r_{ij}\right\rangle$. Since the main contribution to Eq. (\ref{eq7}) 
stems from within the screening length $\lambda$, which in turn is comparable to the 
molecular persistence length \cite{Tin97}, the chain may be treated as rigid such that the 
distance of beads $i,j$ simplifies to $r_{ij}=\left\vert i-j\right\vert d$. Replacing moreover the 
double sum by integrals over $i$ and $j$, we obtain 
\begin{equation}
\mathbf{u}=\frac{\zeta_1 ^{2}}{3\eta }\left( 1+\chi _{\epsilon }\right) 
\mathbf{\nabla }\epsilon +\frac{2\epsilon \zeta_1 }{3\eta }(1+\chi _{S})%
\mathbf{E}_{T},  
\label{eq10}
\end{equation}
where the quantities  $\chi _{\epsilon}$ and $\chi _{S}$ account for hydrodynamic 
interactions (see Fig. 2.) With $\chi_\epsilon=0=\chi_S$ one has the explicit expression for the 
monomer velocity $\mathbf{u}_1$ defined in (\ref{eq3}). 

\begin{figure}[tbp]
\includegraphics[width=\columnwidth]{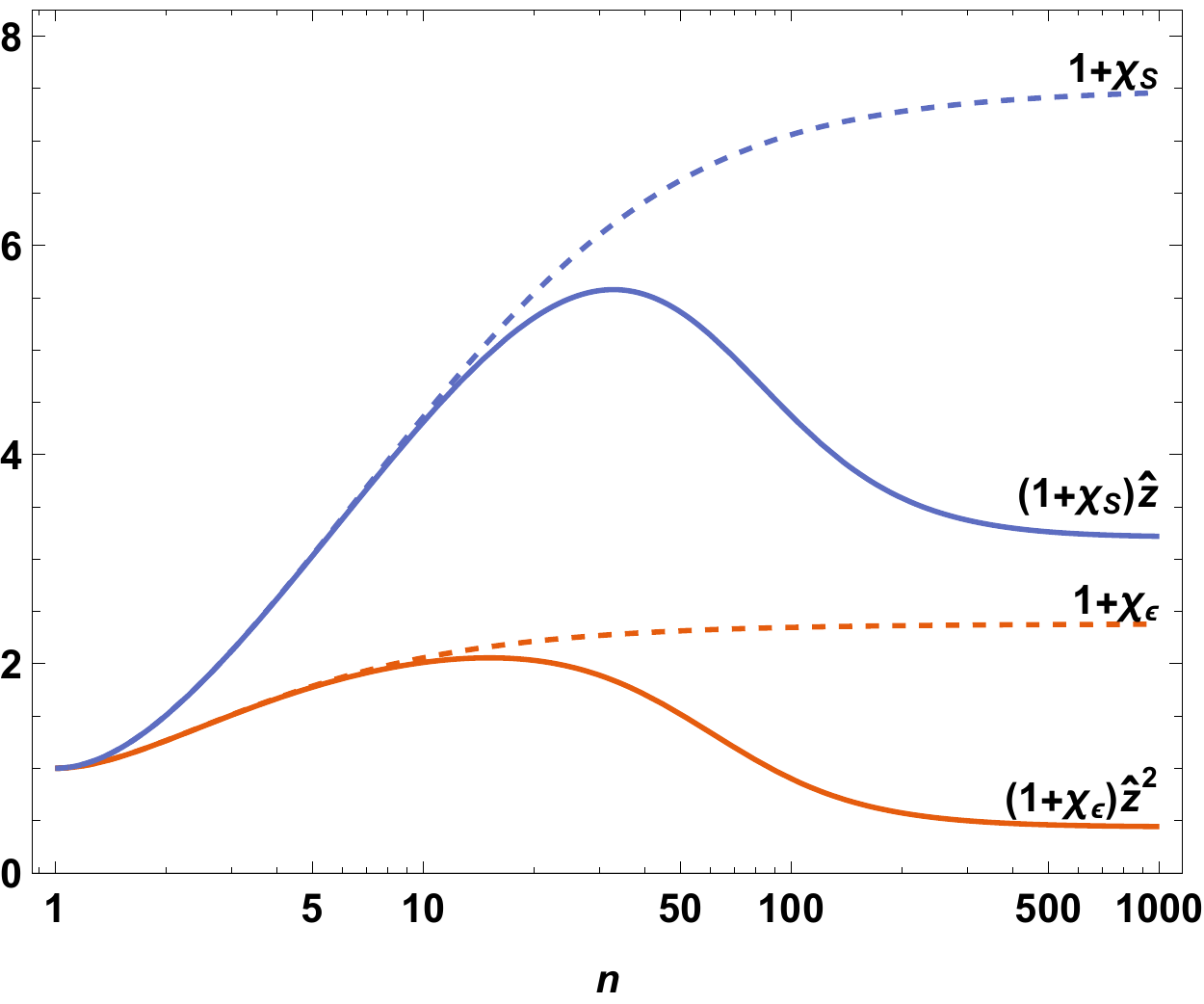}
\caption{Molecular-weight dependence of the two contributions to $D_T$. Both hydrodynamic 
factors  $\protect\chi_\epsilon$ and $\chi _{S}$ increase with $n$, albeit with differnt amplitudes, 
as shown by the dashed lines; the parameters are the monomer distance $d=3\text{\AA }$ and 
the Debye length $\lambda =5$ nm. Counterion condensation results in the factors $\hat{z}^2$ and 
$\hat{z}$, which significantly reduce the mobility (solid lines), according to (\ref{eq11}) with 
$\beta_n=(n^2-1)n_0^{-2}$ and $n_0=80$.}
\end{figure}

The hydrodynamic correction factor for motion driven by the permittivity gradient  
reads 
\begin{equation}
\chi _{\epsilon }=\frac{a^{2}}{d^{2}}
\left( ( 1+2n\hat{d}) \frac{E_{2\hat{d}} - E_{2n\hat{d}} }{n}+e^{-2\hat{d}}-\frac{e^{-2n\hat{d}}}{n}\right) ,
\label{eq10a}
\end{equation}
with the shorthand notation $E_x={\rm Ei}(-x)$ for the exponential integral function, 
and $\hat{d}=d/\lambda$ for the ratio of the monomer length and the Debye length.  
For the Seebeck term we find
\begin{equation}
\chi _{S}=\frac{2a}{d}\left( E_{n\hat d} -E_{\hat d} +\frac{e^{-n\hat{d}}-e^{-\hat{d}}}{n\hat{d}}\right) ,
\end{equation}
The factor $2$ in the exponential and Ei functions in $\chi_\epsilon$ arises from the screening 
factor of the force density, $E^2\propto e^{-2r/\lambda}$, whereas the factors in $\chi_S$  
are related to the decay of the screening cloud, $\rho\propto e^{-r/\lambda}$.
Fig. 2 shows $\chi_\epsilon$ and $\chi_S$ as a function of the molecular weight. Both vanish 
for monomers, $n=1$, whereas for long  molecules they tend toward the constants 
$\chi_\epsilon^\infty=(a/d)^2(2\hat{d}E_{2\hat{d}}+e^{-2\hat{d}})$ and $\chi_S^\infty=-2(a/d) E_{\hat{d}}$. 
Note that $\chi_S$ is identical to the hydrodynamic correction of electrophoresis \cite{Mut94}. 
Flexible molecules, such as the bead-spring model, would result in cumbersome formal expressions 
without changing the qualitative features. For example, evaluating the configuration average with a 
power law for the bead-bead distance, $\left<r_{ij}^2\right>\propto|i-j|^{2\nu}$, results in 
an incomplete Gamma function $\Gamma(\nu^{-1},nd/\lambda)$ instead of an exponential integral; 
the numerical evaluation of the mean velocity (\ref{eq7}) hardly differs from the above correction factors. 

{\it Counterion condensation.--} 
A polyelectrolyte carries a line charge $e/d$. If the bead spacing $d$ is larger than the Bjerrum 
length $l_B \approx 7$\AA, Debye-H\"{u}ckel approximation is valid even for long chains, and 
the electrostatic potential reads $\sum_i\psi_i(\mathbf{r}-\mathbf{r}_i)$. 
Yet in the opposite case $d<l_B$, which is relevant for DNA, this linear superposition ceases 
to be valid as $n$ increases. Because of the strong Coulomb interaction, the counterions 
partly condense onto the polymer until its linear charge density is reduced to the critical value 
$e/l_B$ \cite{Man81}. The remaining free counterions are well described by 
Debye-H\"{u}ckel theory.

In a mean-field model, counterion condensation is described by an effective valency 
\begin{equation}
\hat{z}=\xi ^{-1}+\frac{1-\xi ^{-1}}{1+\beta _{n}},  
\label{eq11}
\end{equation}
where $\xi =l_{B}/d>1$ is the Manning parameter. For a monomer the quantity $\beta _{1}$
vanishes, and one has  $\hat{z}=1$. For long chains, $\beta_n$  tends to infinity, thus resulting 
in $\hat{z}=\xi ^{-1}$ and reducing the charge density to its critical value $\hat{z}e/d=e/l_B$ 
\cite{Man81}. The progressive condensation of the counterions on the chain, and the 
dependence of $\beta _{n}$ on $n$ and $\lambda$, constitute an intricate problem which is 
beyond the scope of the present paper \cite{Gra10}. Here we use the simple form 
$\beta _{n}=(n^{2}-1)n_0^{-2}$ which, with $n_0=80$, fits rather well 
the experimental data. Note that this model does not depend on the electrolyte strength.

\begin{figure}[tbp]
\includegraphics[width=\columnwidth]{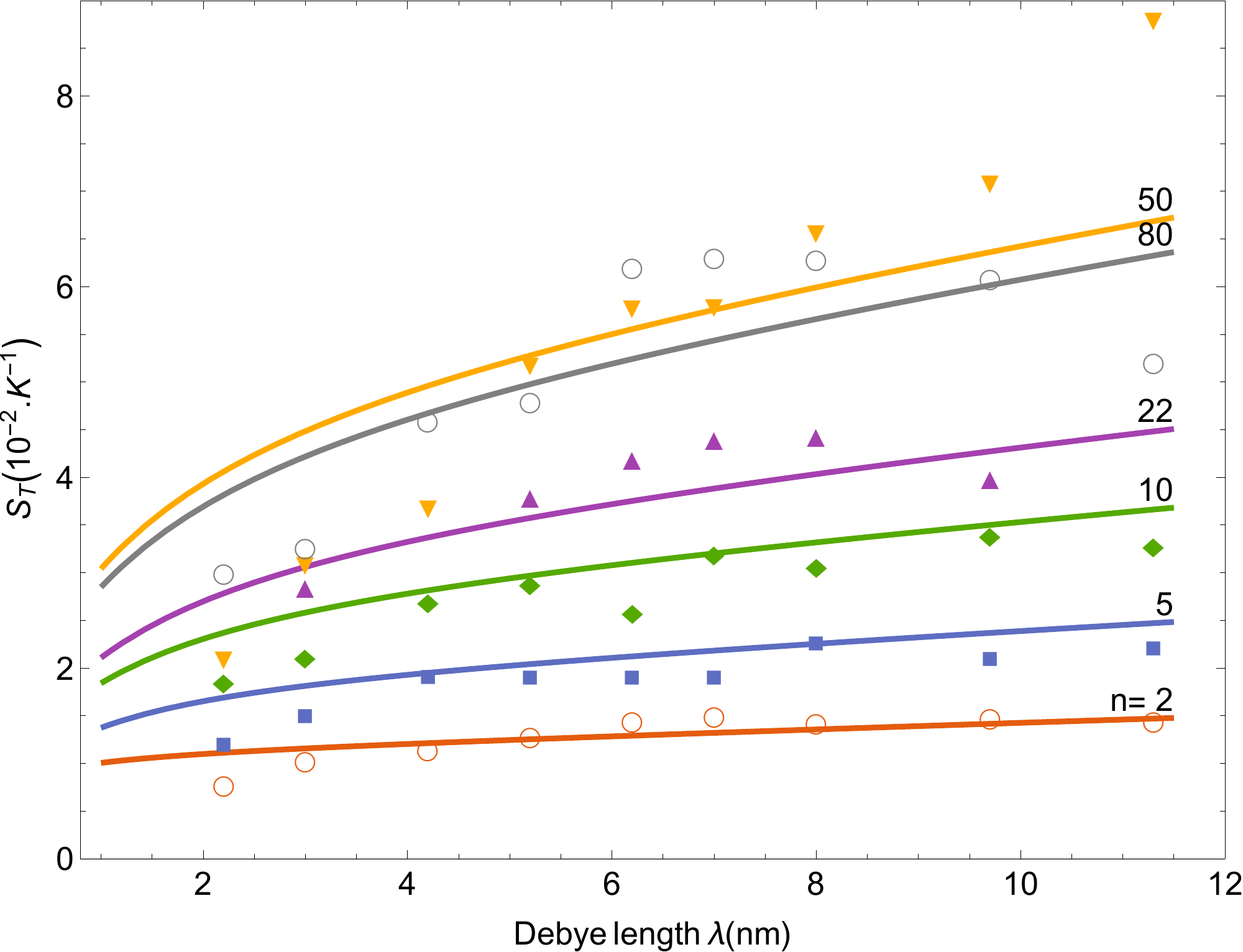}
\caption{The Soret coefficient $S_{T}$ as a function of the Debye length $\lambda$ for DNA of 
different length $n$. The data points, measured at 15$^\circ$ C, are taken from Ref. \cite{Rei14}. 
The theory curves are calculated from $S_T=D_T/D$, where $D_T$ is given by Eq. (\ref{eq12}) 
with $\hat{S}=0$, $a=4.25$\AA,  and $d=3$\AA. The values of the diffusion coefficient $D$ for $n=5,...,50$ 
are obtained from experiment\cite{Rei14}, those for $n=2$ and 80 are extrapolated 
from the previous and from the power law for long molecules \cite{Sim12}.}
\label{fig:ST}
\end{figure}

{\it Phoretic coefficients.--}
The thermophoretic mobility is defined through the drift velocity $u=-D_T\nabla T$ in a temperature 
gradient. From (\ref{eq10}) we find 
\begin{equation}
D_{T}=\frac{k_{B}}{12\pi \eta a}
      \left( \frac{l_{B}}{a} \hat{z}^2 (1+\chi _{\epsilon }) \tau  + 2\hat{z} ( 1+\chi _S) \hat{S} \right),  
\label{eq12}
\end{equation}
with the parameter $\tau=-d\ln \epsilon/d\ln T \approx 1.4$ which arises from the permittivity 
gradient, and the dimensionless Seebeck coefficient $\hat{S}=S(e/k_{B})$. For monomers the 
mobility is independent of the Debye length, whereas for longer chains, the correction factors 
give rise to complex dependencies on $\lambda$ and $n$. In Fig. 2 we plot the 
two contributions to $D_T$ as a function of $n$. The initial increase results from hydrodynamic 
interactions (dashed lines), whereas the decrease at larger $n$ is due to counterion condensation 
(solid lines). Both factors reach a finite value at large $n$; for typical parameters of DNA in 
a weak electrolyte, the permittivity term shows an overall decrease, 
$\hat{z}^2 (1+\chi _{\epsilon}^\infty)<1$, whereas the Seebeck term is enhanced, 
$\hat{z} (1+\chi _S^\infty)>1$. 

The stationary DNA concentration $c$ is achieved when thermophoretic drift 
and gradient diffusion with coefficient $D$ cancel each other, $cu-D\nabla c=0$. This 
``Soret equilibrium'' is usually written in the form $\nabla c+cS_T  \nabla T=0$, since 
experiments probe the Soret coefficient $S_T=D_T/D$  rather than the mobility $D_T$. 
In Fig. \ref{fig:ST} we compare our theory with Soret data for single-stranded DNA as a function 
of the Debye length $\lambda$, taken from Ref. \cite{Rei14}. The theoretical 
curves are calculated with (\ref{eq12}) and a simple model for the measured diffusion 
coefficient $D$, as described in \cite{Supp}.
The best agreement with the data is obtained when retaining in (\ref{eq12}) the 
permittivity-gradient term only, that is, for zero Seebeck coefficient, $\hat{S}=0$. 
The increase of  $S_T$ with the Debye length arises mainly from the hydrodynamic 
correction $\chi_\epsilon$. For short chains, $n<30$, the variation with $n$ is of purely 
hydrodynamic origin, whereas for larger chains counterion condensation plays an 
important role, as is clear from Fig. 2.  

\begin{figure}[tbp]
\includegraphics[width=\columnwidth]{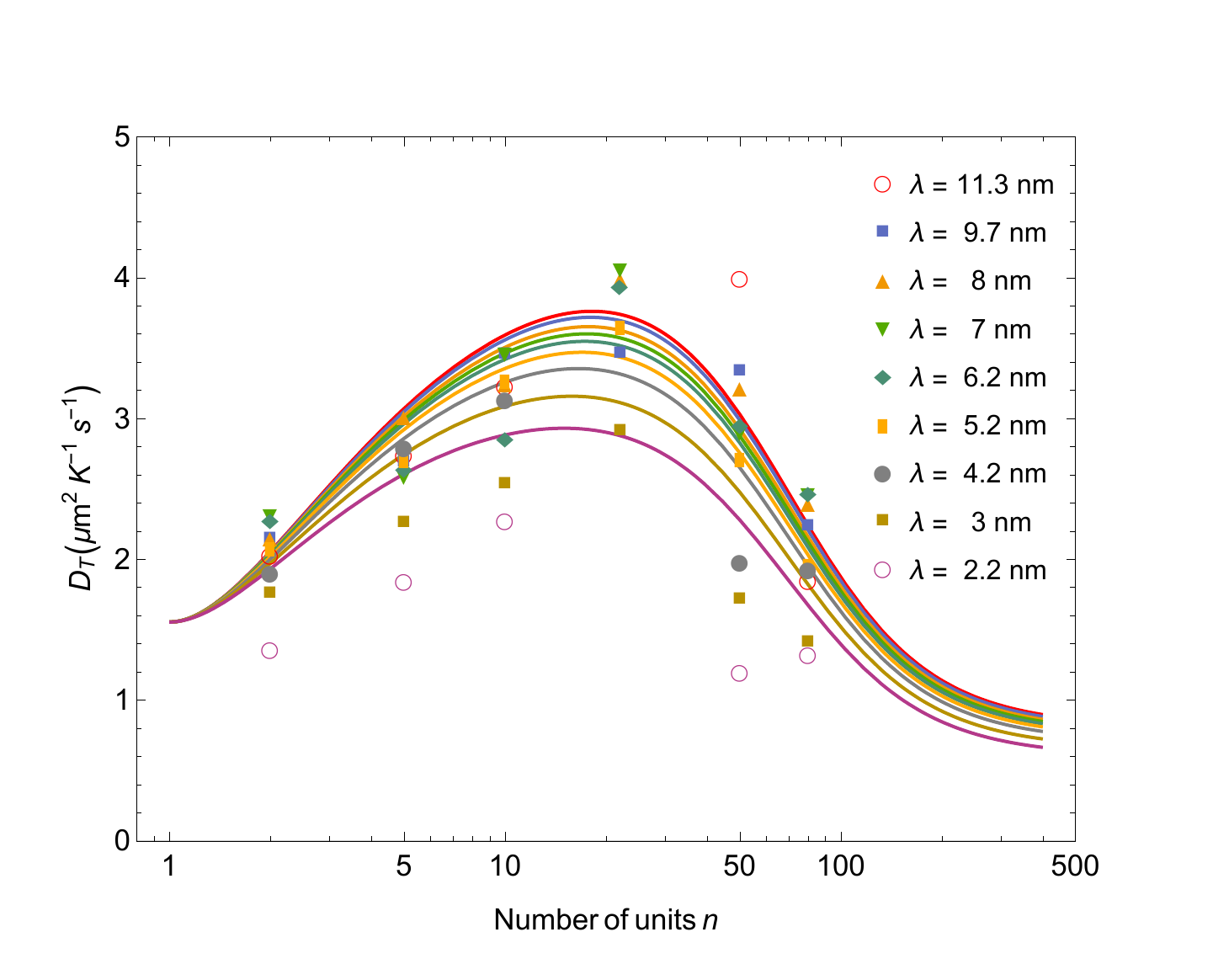}
\caption{Thermophoretic mobility $D_{T}$ as a function of the molecular length $n$ for various 
values of the Debye length $\lambda$. The full curves are calculated from Eq. (\ref{eq12}) 
with $\hat{S}=0$, $a=4.25$\AA\ and $d=3$\AA. The data points give $D_T=DS_T$, with 
$S_T$ and the hydrodynamic radius from Ref. \cite{Rei14}; for details see \cite{Supp}.}
\label{fig:DT}
\end{figure}

In order to clearly display the effect of hydrodynamic interactions, we plot in Fig. \ref{fig:DT} 
the thermophoretic mobility (\ref{eq12})  as a function of the molecular weight $n$.  The 
experimental points are obtained from $D_T=DS_T$, with measured $S_T$ and $D$  \cite{Rei14} 
as described in \cite{Supp}. The theoretical curves are 
calculated with the permittivity-gradient only ($\hat{S}=0$).  The initial increase 
of the data up to $n=22$  agrees well with the relation (\ref{eq10a}), thus 
providing strong evidence for the role of hydrodynamic interactions. The maximum and the 
subsequent decrease are well described by counterion condensation according to (\ref{eq11}).
Adding a significant thermoelectric contribution would not improve the quality of the fit, quite
 on the contrary. This suggests that the Seebeck field in NaCl solution is small,  confirming a 
previous analysis of Soret data for polystyrene beads \cite{Esl14}. 

\begin{figure}[t]
\includegraphics[width=\columnwidth]{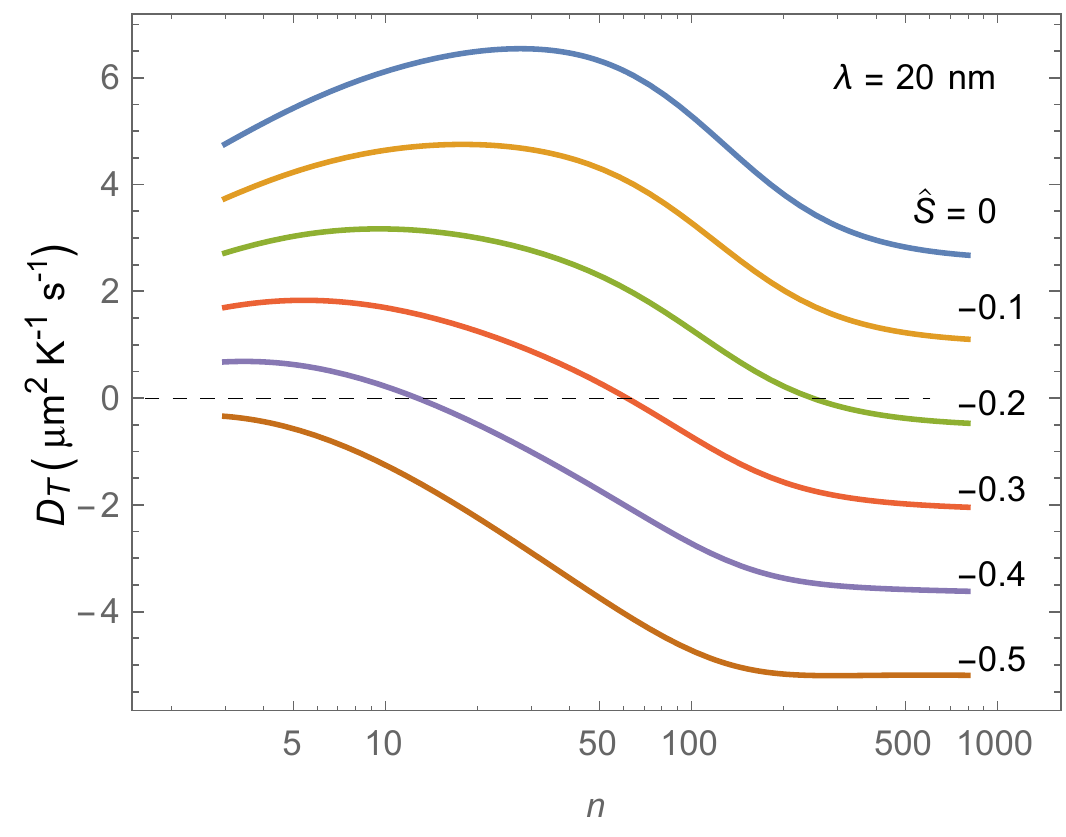}
\caption{Thermophoretic mobility $D_{T}$ as a function of the chain length $n$, for different 
values of the dimensionless Seebeck coefficient $\hat{S}$. For negative  $\hat{S}$ the 
thermoelectric field in (\ref{eq10}) drives the molecules toward the hot, whereas the permittivity 
gradient points toward the cold. Since the latter dominates for short molecules and the latter for 
long ones, $D_T$ changes sign as the $n$ increases. }
\label{fig:n}
\end{figure}  

The electrolyte Seebeck effect was discarded in the above analysis of Soret data in NaCl 
solution. In Fig. \ref{fig:n} we plot the complete mobility $D_T$ as a function of $n$, for 
several values of the dimensionless Seebeck coefficient $\hat{S}$. As the most striking 
feature,  for negative $\hat{S}$ the superposition of the two contributions in (\ref{eq12}) 
may result in a change of sign of the $D_T$. From Fig. 2 it is clear that for short chains, 
the permittivity gradient term prevails, whereas for longer moleculese the Seebeck term 
dominates because of its much larger hydrodynamic factor $\chi_S$. 

The resulting velocity difference could be used for specific accumulation of one component 
at a heated spot, or for separating DNA by molecular size. For example, in an electrolyte 
with $\hat{S}=-0.3$, the permittivity-gradient term dominates for short molecules 
($n<50$) which move to the cold accordingly, whereas longer chains ($n>50$) are driven 
to the hot by the thermoelectric field $E_T$. The stagnation molecular length $n_c$, 
where $D_T=0$, is easily adapted by chosing an appropriate salt mixture. The change of sign 
has been observed for nano-size micelles \cite{Vig10} and micron-size polystyrene beads 
\cite{Esl14} in mixed electrolytes NaOH$_x$Cl$_{1-x}$; the values of $\hat{S}$ used in Fig. 
\ref{fig:n} are realized by with $0.1<x<0.4$.   

{\it Conclusion. --} We briefly summarize our main results on DNA thermophoresis. 
We find that DNA thermophoresis is rather sensitive to hydrodynamic interactions. With 
increasing chain length $n$, the mobility is enhanced due to mutual advection of the repeat units, 
then passes through a maxium, and finally decreases below the monomer value due to counterion 
condensation. Comparison with measured data in Fig. 4, provides strong evidence that the 
molecular-weight dependence arises from the interplay of hydrodynamic interactions and 
non-linear charge effects. The interplay between the dominant driving forces, that is, the permittivity gradient and a 
thermoellectric field with negative Seebeck coefficient, results in a change of sign as a function of 
$n$: Short molecules move to the cold, and long ones to the hot.

The authors acknowledge funding by the French National Research Agency through
contract ANR-13-IS04-0003.

\end{document}